\documentclass[
 reprint,
preprintnumbers,
 amsmath,amssymb,
 aps]{revtex4-1}
\usepackage{cancel}
\usepackage{xcolor}

\usepackage[normalem]{ulem}
\usepackage[colorlinks,pdfusetitle]{hyperref}
\usepackage{multirow}
\usepackage{graphicx}% Include figure files
\usepackage{dcolumn}% Align table columns on decimal point
\usepackage{bm}% bold math
\usepackage{tikz}

\begin{document}

\preprint{FERMILAB-PUB-24-0117-T}

\title{
A Mass Ordering Sum Rule for the Neutrino Disappearance Channels\\ in T2K, NOvA and JUNO 
}

%\author{Hiroshi Nunokawa}
%\email{nunokawa@puc-rio.br; \# orcid: 0000-0002-3369-0840}
%\affiliation{%
 %Departamento 
 %Dept. de F\'isica, Pontif\'icia Universidade Cat\'olica do Rio de Janeiro, 
 %C.P. 38071, 
 %22452-970, Rio de Janeiro, Brazil }
 %
 \author{Stephen J. Parke}
\email{parke@fnal.gov; \# orcid: 0000-0003-2028-6782}
\affiliation{Theoretical Physics Department, Fermi National Accelerator Laboratory, Batavia, IL 60510, USA}
 \author{Renata Zukanovich Funchal}%
 \email{zukanov@if.usp.br; \# orcid: 0000-0001-6749-0022}
\affiliation{%
 Instituto de F\'isica, Universidade de S\~ao Paulo, 
 %C.P. 66.318, 
 05315-970 S\~ao Paulo, Brazil
}

\date{April 12, 2024}

\begin{abstract}
We revisit a method for determining the neutrino mass ordering by using precision measurements of the atmospheric $\Delta m^2$'s in both electron neutrino and muon neutrino disappearance channels, proposed by the authors in 2005~\cite{Nunokawa:2005nx}.  The mass ordering is a very important outstanding question for our understanding of the elusive neutrino and determination of the mass ordering has consequences for other neutrino experiments.  The JUNO reactor experiment will start data taking this year, and the precision of the atmospheric $\Delta m^2$'s from electron anti-neutrino measurements will improve by a factor  of three from Daya Bay's 2.4\% to 0.8\% within a year. This measurement, when combined with the atmospheric $\Delta m^2$'s measurements from T2K and NOvA for muon neutrino disappearance, will contribute substantially to the $\Delta \chi^2$ between the two  remaining neutrino mass orderings.  In this paper we derive a mass ordering sum rule that can be used to address the possibility that JUNO's  atmospheric $\Delta m^2$'s measurement,  when combined with other experiments  in particular T2K and NOvA, can determine the neutrino mass ordering  at the 3 $\sigma$ confidence level within one year of operation.  For a confidence level of 5 $\sigma$ in a single experiment we will have to wait until the middle of the next decade when the DUNE experiment is operating.
\end{abstract}

\maketitle

\begin{figure}[!b]
        \vspace*{-5mm}
           \includegraphics[width=0.32\textwidth]{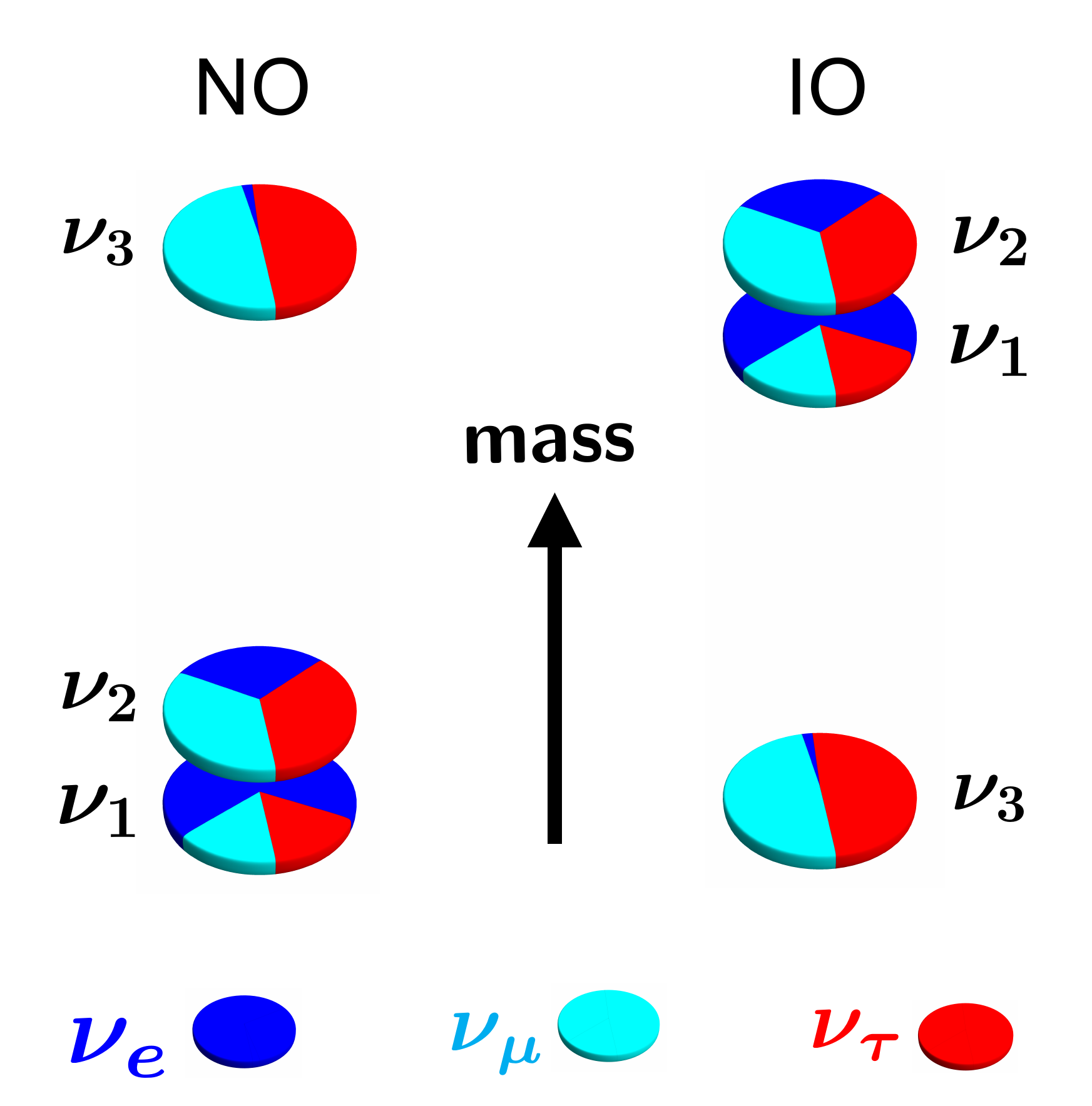}
        \caption{The two remaining mass orderings for the 3 neutrino mass states, from  \cite{Parke:2018shx}. Dark blue is the $\nu_e$ fraction, cyan the $\nu_\mu$ fraction and red the $\nu_\tau$ fraction.  If the mass state with the least fraction of $\nu_e$, labelled $\nu_3$, is at the top of the spectrum this is called the normal ordering (labelled NO) whereas if it is at the bottom of the spectrum it is called the inverted ordering (IO). SNO~\cite{SNO:2011hxd} determined the mass ordering of the other two mass states using  solar neutrinos. The set $(\nu_1, \nu_2)$, usually called the solar pair, has the state with most $\nu_e$, labelled $\nu_1$,  below the other member of the pair, labelled $\nu_2$.         }
            \label{fig:nuMO}
        \end{figure}

\textbf{\textit{Introduction}}  ---
We have known for more than a quarter of century that neutrinos are massive \cite{Super-Kamiokande:1998kpq}  but we still do not know whether the neutrino with the least amount of $\nu_e$, usually labelled $\nu_3$, is at the top or bottom of the neutrino mass spectrum.  This is the neutrino mass ordering question.  The SNO experiment~\cite{SNO:2011hxd} determined that the mass ordering of the other two neutrino mass eigenstates was such that the neutrino with the most $\nu_e$, usually labelled $\nu_1$, was lighter than the other mass state, $\nu_2$ which 
has a smaller $\nu_e$ fraction than $\nu_1$ but a larger $\nu_e$ fraction than $\nu_3$. Thus, the remaining possible mass ordering  for the neutrino mass states is, either $m_1 < m_2 < m_3$ which is known as the normal ordering (NO)  or  $m_3 < m_1 < m_2$ which is known as the inverted ordering (IO), see Fig.   \ref{fig:nuMO}. The mass squared splitting between $\nu_2$ and $\nu_1$ was measured with good precision by the KamLAND experiment to be~\cite{KamLAND:2013rgu}
\begin{align}
\Delta m^2_{21} \equiv m^2_2-m^2_1 \approx  + 7.5 \times 10^{-5} ~ {\rm eV}^2 \, .
\label{eq:dmsq21}
\end{align}
Whereas the magnitude of the mass-squared-splitting between $\nu_3$ and $\nu_1$ has been determined by a number of experiments to be 30 times larger, i.e.
\begin{align}
\Delta m^2_{31} \equiv m^2_3-m^2_1 \approx   \pm \, 2.5 \times 10^{-3} ~ {\rm eV}^2\, , 
\label{eq:dmsq31}
\end{align}
where the ambiguity in the sign comes from the undetermined mass ordering\footnote{For NO $\Delta m^2_{atm}>0$ and for IO $\Delta m^2_{atm}<0$ for any atmospheric $\Delta m^2$, atm =(ee, $\mu \mu$, 31, 32).}. 
 There exist numerous ways to determine the mass ordering and hence the above sign in the literature~\cite{Dighe:1999bi,Minakata:2000rx,Minakata:2001qm,Lunardini:2001pb,Barger:2001yr,Barger:2002rr,Petcov:2001sy,Huber:2002rs,Dighe:2003jg,Dighe:2003be,
Lunardini:2003eh,Mena:2004sa,Barger:2005it,Nunokawa:2005nx}. 
Nevertheless, the use of matter effects in neutrino oscillations have been guiding most of the experimental efforts. 
The long-baseline accelerator neutrino experiments NOvA~\cite{NOvA:2023iam} and T2K~\cite{T2K:2023mcm} 
  as well  as the atmospheric neutrinos experiments 
  Super-Kamiokande~\cite{Super-Kamiokande:2023ahc} and Ice-Cube~\cite{IceCube:2014flw} operate in a regime where 
  neutrino oscillations are mostly 
  driven by  the higher  mass-squared-splitting and matter effects are significant in the $\nu_\mu \to \nu_e$ (and antineutrino) appearance probabilities.
  They are responsible for the current status of our understanding of the mass ordering question.
  At present T2K and NOvA, individually,  have a slight preference for NO, while the combined fit, flips this preference to IO~\cite{Kelly:2020fkv,Esteban:2020cvm}, although the preference is weak.
  Ice-Cube has no preference for either NO or IO.
  On the other hand,  Super-Kamiokande  data seems to favor NO by 92.3\% CL even though their Monte Carlo simulation indicates they should not 
  be able to  discriminate the ordering better than  $\sim 80$\% CL, see \cite{Super-Kamiokande:2023ahc}.  It is the statistical weight of their data,  corresponding to an exposure of 364.8 kiloton-years, that makes the final global data fit prefer NO~\cite{Esteban:2020cvm}.
 So, at the current time, we do not have 3 $\sigma$ or more preference for one ordering over the other.

 In this article we revisit one way proposed by the authors of this paper  \cite{Nunokawa:2005nx} which requires precision measurements of 
$|\Delta m^2_{31}| $ or $|\Delta m^2_{32}|$ by both $\nu_e$/$\bar \nu_e$ and $\nu_\mu$/$\bar \nu_\mu$ disappearance experiments.
The reason for writing this paper now  is that within the next year the JUNO experiment \cite{JUNO:2015zny, JUNO:2022mxj} is expected to  improve the precision of these measurements for
$\bar \nu_e$ disappearance by a factor of approximately 3, i.e. from Daya Bay's 2.4\,\% to better than 0.8\,\%, a very significant improvement. 

\begin{center}
        \begin{figure}[!b]
           \includegraphics[width=0.5\textwidth]{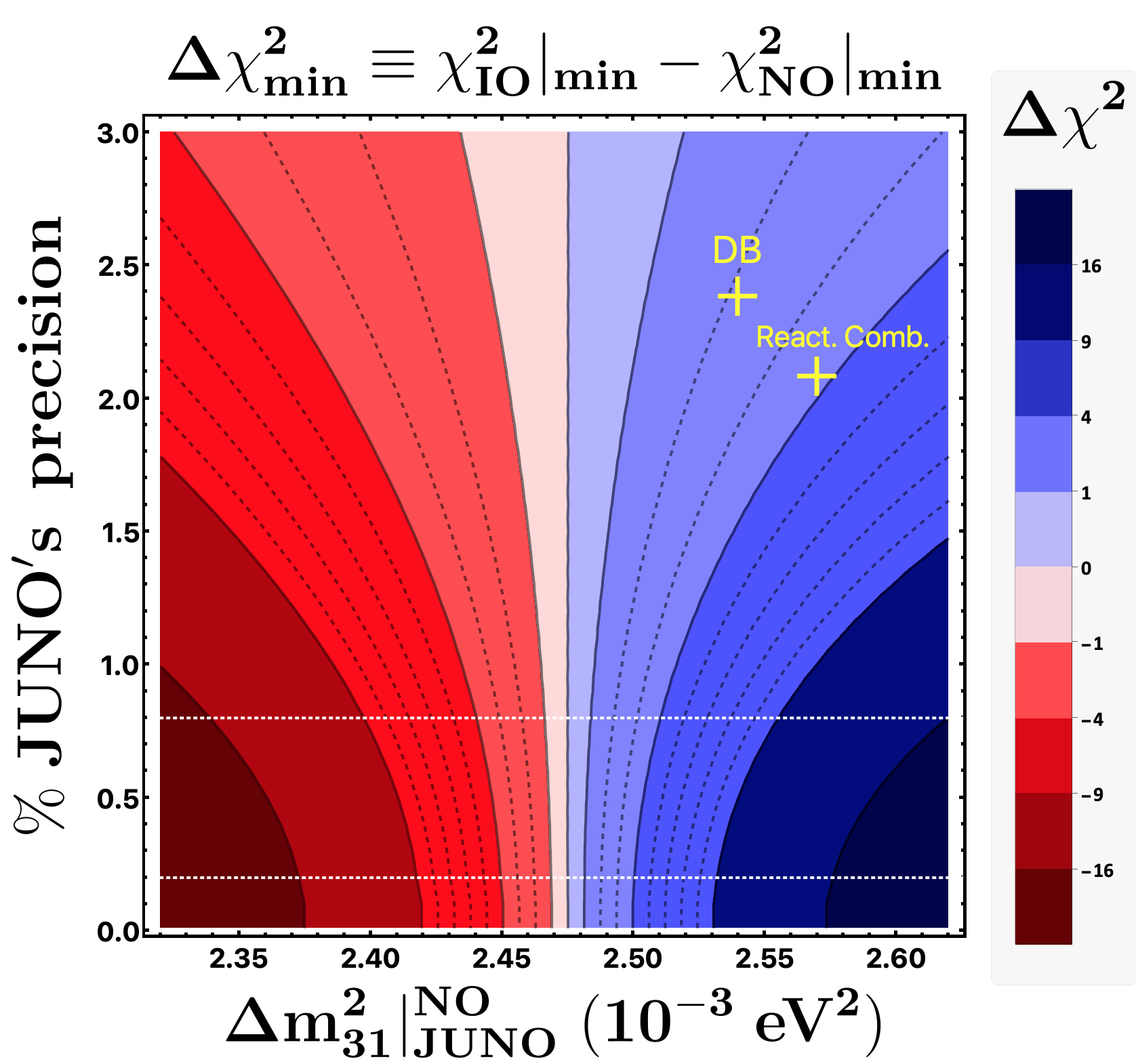}
        \caption{ Iso-contours of $\Delta \chi^2_{\rm min} \equiv \chi^2_{\rm IO}\vert_{\rm min}- \chi^2_{\rm NO}\vert_{\rm min}$ in the plane of JUNO's
$(\Delta m^2_{31}|^{\rm NO} \otimes \sigma )$ fit to the far detector spectrum assuming normal ordering (NO). The precision is expressed as a \% of $\Delta m^2_{31}|^{\rm NO}$. 
 The blue region favors NO whereas the red region favors IO.            The current values from the Daya Bay (DB) experiment and from  Daya Bay and RENO  combined
         according to NuFIT   (React. Comb.)  are also shown. Table \ref{tab:dmsq} can be used to translate the horizontal axis to any other $\Delta m^2_{\rm atm}$. 
         The white dashed lines mark the precision achievable by JUNO after 100 days 
        (0.8\%)  and  6 years (0.2\%).}      
            \label{fig:JUNOplot}
        \end{figure}
    \end{center}

There are a number of papers, such as \cite{Blennow:2012gj, Li:2013zyd, Cabrera:2020ksc}, that use the method first presented in \cite{Nunokawa:2005nx} to address the mass ordering question.
Here, we revisit the combination of long baseline experiments and reactor experiments by simplifying and extend all of these analyses. First we derive a mass ordering sum rule, eq.~\ref{eq:MOsum}, for the electron and muon neutrino disappearance channels.
Then we use NuFIT's combined T2K and NOvA analysis to generate a contour plot of the $\Delta \chi^2$ between the mass orderings in the plane of JUNO's measurement of $\Delta m^2_{\rm 31}[\rm NO]$ and its precision, Fig. \ref{fig:JUNOplot}. This plot allows the reader to immediately estimate the  $\Delta \chi^2$ between the two mass orderings as soon as JUNO presents their $\Delta m^2_{\rm 31}\vert^{\rm NO}$ measurement with its corresponding precision.  Neither the sum rule or the contour plot has appeared elsewhere.

In \cite{Nunokawa:2005nx}, with more details available in \cite{Parke:2016joa}, we have shown 
that the effective atmospheric $\Delta m^2$ ($\Delta m^2_{\rm atm}$) for $\nu_e$ and $\bar{\nu}_e$ disappearance at a baseline divided by neutrino energy of 0.5 km/GeV, in vacuum, is given by\footnote{We use the PDG conventions for the neutrino mixing matrix.}
\begin{equation}
\Delta m^2_{\rm ee} = \Delta m^2_{31} \cos^2 \theta_{12} + \Delta m^2_{32} \sin^2 \theta_{12}\, .
\label{eq:dmsqee}
\end{equation}
This is the only $\Delta m^2$ measured by Daya Bay \cite{DayaBay:2016ggj} or RENO \cite{RENO:2018dro}  without additional information from other experiments. The final precision obtained by Daya Bay (DB) was  2.4\%, see \cite{DayaBay:2022orm}:
\begin{align}
|\Delta m^2_{\rm ee}|_{\rm DB}= 2.519 \pm 0.060 ~~{\rm eV}^2 \, .
\label{eq:dmsqeeDB}
\end{align}
Since the  Daya Bay experiment is insensitive to the neutrino mass ordering, the magnitude of $\Delta m^2_{\rm ee}$ is the same for both orderings and the sign is undetermined. Matter effects on the magnitude of  $\Delta m^2_{\rm ee}$ are smaller than one tenth of one per cent level, see~\cite{Khan:2019doq},  and are therefore much smaller than the measurement uncertainties. 

Similarly, for  $\nu_\mu$ (and $\bar{\nu}_\mu$) disappearance, in vacuum,  $\Delta m^2_{\mu \mu}$ 
is given by 
\begin{align}
\Delta m^2_{\rm \mu \mu} \approx   & \; \Delta m^2_{31} \sin^2 \theta_{12} + \Delta m^2_{32} \cos^2 \theta_{12} \notag \\
&+ \sin \theta_{13} \cos \delta ~ \Delta m^2_{21}\, ,
\label{eq:dmsqmm}
\end{align}
where $\delta$ is the CP phase and  we have set $\sin 2\theta_{12} \tan \theta_{23} \approx 1$ for simplicity in the coefficient of the $\cos \delta$ term.  For the current long baseline experiments such as T2K and NOvA, even though matter effects are significant for the appearance channels,  it was recently shown by Denton and Parke, in~\cite{Denton:2024thm}, that for the $\nu_\mu$ disappearance channels matter effects are at the tenths of one per cent level. 
Therefore the $\nu_\mu$ disappearance channels are effectively in vacuum as there is a cancellation between the matter effects in $\nu_\mu \rightarrow \nu_e$ and in 
$\nu_\mu \rightarrow \nu_\tau$ for the long baseline experiments T2K and NOvA.
As a result these disappearance channels  are also insensitive to the neutrino mass ordering.
Similar to Daya Bay and RENO, $|\Delta m^2_{\rm \mu \mu} |$ is the only $\Delta m^2$ that is measurable in T2K and NOvA without information from other experiments.

For NO  $ \Delta m^2_{31} > \Delta m^2_{32}$ whereas for IO
$ |\Delta m^2_{31}| < |\Delta m^2_{32}|$, therefore, since $\sin^2\theta_{12} \approx 0.3$ we can determine the mass ordering by comparing  the magnitude of $|\Delta m^2_{\rm ee} |$  to $| \Delta m^2_{\rm \mu \mu} |$:
\begin{align}
&|\Delta m^2_{\rm ee}|
> |\Delta m^2_{\rm \mu \mu}|  \text{ for NO} \notag \\
& \hspace{1.5cm} {\rm whereas}~ |\Delta m^2_{\rm ee}| < |\Delta m^2_{\rm \mu \mu} | ~\text{for IO},
\end{align}
although the size of the difference between $|\Delta m^2_{\rm ee} |$ and 
$| \Delta m^2_{\rm \mu \mu} |$ is at the couple of percent level and therefore precise measurements are required~\footnote{If T2K and NOvA reported the $ |\Delta m^2_{\rm \mu \mu} | $ fits to their disappearance only data, this comparison could be made more directly without any information needed on $\cos \delta$.}.  Fortunately we are about to enter an era where very precise measurements will be  made for 
$\bar{\nu}_e$ disappearance by the JUNO experiment.

%\vspace{4cm}

 \begin{widetext}
 The above observations can be converted into a  {\it ``mass ordering sum rule for the neutrino disappearance channels'' }:
 \begin{align}
( \Delta m^2_{31}|^{\rm NO}_{\mu ~\rm disp} - \Delta m^2_{31}|^{\rm NO}_{e ~\rm disp}) 
~~ + ~~ (| \Delta m^2_{32}|^{\rm IO}_{e ~\rm disp} - |\Delta m^2_{32}|^{\rm IO}_{\mu ~\rm disp}) 
~~ = & ~~(2 \cos 2 \theta_{12}- 2 \sin \theta_{13} \, \widehat{\cos \delta} )\Delta m^2_{21}\, , \label{eq:MOsum}
 \end{align}
 where the RHS can also be written as $(2.4 -0.9  \, \widehat{\cos \delta})\% ~ |\Delta m^2_{\rm atm}| $.
 The subscript ``$\mu$ disp" means the results from $\nu_\mu$ disappearance measurements in T2K and NOvA, whereas ``e disp'' means the result from $\bar{\nu}_e$ disappearance experiments such as Daya Bay and RENO.  The symbol $ \widehat{\cos \delta}$ is the average $\cos \delta$ for the NO and IO fits.    If one changes which  $\Delta m^2$ one uses for both experiments for a given mass ordering, the RHS of this sum rule is unchanged.  For detailed derivation of this sum rule see  the appendix/supplemental material.\\
  \end{widetext}
 
 From this sum rule it is clear that if NO is Nature's choice then the first term will be zero, within measurement uncertainties, and if IO is Nature's choice the second term will be zero but in both scenarios  the sum of the two must add up to the RHS, independent of the mass ordering.  Consequently,  if NO is Nature's ordering,  the measurements of the $\Delta m^2_{32}$'s assuming IO will not align between $\nu_e$ and $\nu_\mu$ disappearance  within $\approx$ 2.4\%  and similar for the IO ordering. The measurement uncertainties of the experiments Daya Bay, T2K and NOvA are now small enough that this method is already contributing to the global fits on the neutrino mass ordering.  This can be perceived in the  latest  \href{http://www.nu-fit.org/sites/default/files/v52.fig-chisq-dma.pdf}{NuFIT  figure on the synergies for the $\Delta m^2_{3\ell}$'s}, see \cite{NuFIT2022}, one can observe a preference for NO although the precision of the current measurements is not sufficient for a 3 $\sigma$ determination of the neutrino mass ordering.
 
 The question of immediate current interest is how will the precision measurements of the  $\Delta m^2_{3i}$'s  by JUNO affect the determination of the mass ordering as JUNO measurements are expected to have an uncertainty smaller than one third of Daya Bay's.  
 This measurement   is expected to come very quickly after JUNO turns on, most likely in the first year of operation. \\

\noindent \textbf{\textit{ Reactor Measurement ($\bar{\nu}_e$ Disappearance)}} ---
JUNO is a medium baseline ($\sim50$ km) high precision reactor antineutrino oscillation experiment  
aiming to determining the neutrino mass ordering by a careful measurement of the 
$\bar \nu_e$ energy spectrum using an idea first proposed in~\cite{Petcov:2001sy} 
and further investigated in~\cite{Choubey:2003qx,Bilenky:2017rzu}.
It was shown in \cite{Minakata:2007tn} that medium baseline reactor 
experiments can, in principle, determine the ordering by precisely measuring  the effective combination 
$\Delta m^2_{\rm ee}$  and the sign of a phase ($\pm \Phi_\odot$; $+$ for NO, $-$ for IO). 
This is a very challenging measurement due to various 
systematic effects (energy resolution, non-linear detector response etc.) 
that have to be tamed and understood to a very high level. 
The JUNO collaboration claimed, in their 2015 paper \cite{JUNO:2015zny}, that it will take 6 years to determine the mass ordering at 3 $\sigma$, although more recent papers, see e.g.  \cite{Forero:2021lax}, have questioned that claim\footnote{Two 4.6 GW$_{th}$ were not built  at Taishan and the best fit values of the relevant parameters $\sin^2 \theta_{13}$ and $\Delta m^2_{atm}$ by Daya Bay, both moved in the unfavorable directions, see Fig 8. of  \cite{Forero:2021lax}, compared to what was used in  \cite{JUNO:2015zny}.  } . JUNO's recent update \cite{JUNO:2022mxj} does not contain  an update  on their expected mass ordering sensitivity.\\

On the other hand, JUNO is expected to reach, after a few months of operation, 
 and much sooner than  they can start to be sensitive to the mass ordering, 
unprecedented sub-percent precision  on the determination of  
$|\Delta m^2_{\rm atm}|$ where $|\Delta m^2_{\rm atm}|$ could be any one of  $\vert \Delta m^2_{\rm ee}\vert$,  or $\vert \Delta m^2_{31}\vert$, or $\vert \Delta m^2_{32}\vert$ depending on the experiment's analysis choice for both mass orderings but all are related to one another.
JUNO claims that after 100 days 
of data taking they will be able to determine
$\vert \Delta m^2_{\rm atm}\vert$ at  $0.8\%$ precision and will continue to improve ultimately reaching 0.2\% precision, see  \cite{JUNO:2022mxj}.
In  contrast, the expected $\Delta \chi^2$ between the NO and IO  in JUNO's fits is expected to grow quite slowly, at no more than 1.5 units per year.  \\

It was shown in \cite{Forero:2021lax}, that due to the phase  advance (NO)  or retardation (IO) the best fits to the spectrum at the far JUNO (JU) detector will give a
 $\vert \Delta m^2_{\rm ee}\vert$  for IO which  is 0.7\% larger than 
$\vert \Delta m^2_{\rm ee}\vert$ fit for NO, i.e. 
\begin{align}
|\Delta m^2_{\rm ee}|^{\rm IO}_{\rm JU} &=
 \Delta m^2_{\rm ee}|^{ \rm NO}_{\rm JU} +1.8 \times 10^{-5} ~{\rm eV}^2  \, .
 \label{eq:Jee}
\end{align}
Note this shift is a fraction of $\Delta m^2_{21}$ and comes from the fact that in the IO the atmospheric oscillations are retarded with respect to the NO and therefore $\Delta m^2_{ee}$ for IO is slightly larger than that for NO to compensate for this phase shift.
We have used that  $0.007 \times \vert \Delta m^2_{\rm ee}\vert \approx 0.018 \times 10^{-3} {\rm eV}^2$ and the value of $\Delta m^2_{21}$ is given by eq. \ref{eq:dmsq21}.  Using eq. \ref{eq:dmsqee} for each mass ordering we can relate this result for any of the other possible $\vert \Delta m^2_{\rm atm}\vert$ as given in Table \ref{tab:dmsq}.

\begin{table}[t]
{\large
\begin{tabular}{|c||c|c|c|}
\hline
 $\Delta (\Delta m^2)$ & NO ee & NO 31 & NO 32\\
\hline \hline
IO ee & 1.8 & -0.5 & 6.9  \\
IO 31 & -0.5 & -2.7 & 4.7 \\
IO 32 & 6.9 & 4.7 & 12.1\\
\hline
\end{tabular}
}
\caption{This Table relates the different $\Delta m^2_{\rm atm}$'s for the best fits of the JUNO experiment, such that each entry gives the difference between $\Delta (\Delta m^2) 
\equiv  |\Delta m^2_{ij}|^{\rm IO} -\Delta m^2_{kl}\vert^{\rm NO}$  in units of $10^{-5}$ eV$^2$ where (ij)  and (kl) are (ee, 31, 32). A value of 2.5 in this table represents a 1.0\,\% difference. The difference between $\vert\Delta m^2_{32}\vert^{\rm IO}$ and
$ \Delta m^2_{32}\vert^{\rm NO}$ is $\sim 5$\%, much larger than the measurement error expected in JUNO, whereas the difference between $|\Delta m^2_{31}|^{\rm IO}$ and
$ \Delta m^2_{31}\vert^{\rm NO}$ is $\sim 1$\%.}
\label{tab:dmsq}
\end{table}

This modifies the {\it ``mass ordering sum rule for the neutrino disappearance channels'' }, given in eq. \ref{eq:MOsum}, 
 by increasing the RHS by 0.7\%, so that for T2K, NOvA (LBL) and JUNO (JU) we have the following sum rule
\begin{align}
&( \Delta m^2_{31}|^{\rm NO}_{\rm LBL} - \Delta m^2_{31}|^{\rm NO}_{\rm JU}) 
+ (| \Delta m^2_{32}|^{\rm IO}_{\rm JU} - |\Delta m^2_{32}|^{\rm IO}_{\rm LBL})
\notag
\\[2mm]
&\hspace{2cm}  \approx  (3.1 -0.9  \, \widehat{\cos \delta})\% ~ |\Delta m^2_{\rm atm}|  \,. \label{eq:MOsumJ}
 \end{align}
 With enough precision on the measurements of the $\Delta m^2_{\rm atm}$'s between $\nu_e$ and $\nu_\mu$ disappearance, the $\Delta \chi^2$ between the two  mass ordering fits, can contribute significantly to the determination of the mass ordering.

We can describe for JUNO  the $\chi^2$ fit to data to determine 
 $\vert \Delta m^2_{\rm atm}\vert$, as the parabola which will 
 depend on the assumed best fit value
 $ \Delta m^2_{\rm atm}|^{\rm NO}$ 
 or   $\vert \Delta m^2_{\rm atm}\vert^{\rm IO}$ and $\sigma_{\rm JU}$, the precision of the  measurement, 
 which does not depend on the ordering
\begin{align}
\chi^2_{\rm NO} (\Delta m^2, \sigma_{\rm JU})_{\rm JU}
& = \left( \frac{ \Delta m^2- \Delta m^2_{\rm atm}|^{ \rm NO} } {\sigma_{\rm JU}} 
\right) ^2 \\
\chi^2_{\rm IO} (|\Delta m^2|, \sigma_{\rm JU})_{\rm JU}
& = \left( \frac{ |\Delta m^2|-\vert \Delta m^2_{\rm atm} \vert^{ \rm IO} } {\sigma_{\rm JU}} 
\right) ^2
\end{align}
Note that the subscript `atm' can be any one of (ee, 31, 32) and they do not need to be the same for NO and IO. However, the best fit points must be related by the numbers given in Table \ref{tab:dmsq}.
 Different experiments and different global fit groups may make different choices. Here, we will use the choice used by the NuFIT collaboration, that is $\Delta m^2_{31}$ for NO and $\Delta m^2_{32}$ for IO and the relationship that for the JUNO experiment
 \begin{align}
  \vert \Delta m^2_{32}\vert^{\rm IO}_{\rm JU} & =  \Delta m^2_{31} \vert^{\rm NO}_{\rm JU} + 4.7 \times 10^{-5}~{\rm eV}^2\, .
 \label{eq:dmsq32IO31NO}
\end{align}
The physics conclusions will be independent of the (31, 32) arbitrary choices.\\

 \textbf{\textit{Long-baseline Accelerator Measurement ($\nu_\mu/\bar{\nu}_\mu$ Disappearance)}} ---
T2K  is a long-baseline ($\sim 295$ km)
accelerator neutrino oscillation experiment in Japan
that has collected a total of $1.97 \times 10^{21}$ and 
 $1.63 \times 10^{21}$ protons on target in neutrino and 
 antineutrino modes, respectively. 
 Similarly, NOvA is a long-baseline ($\sim 810$ km) 
 accelerator neutrino oscillation experiment in the US
that also  has collected  $1.36\times 10^{21}$ and 
 $1.25\times 10^{21}$ protons on target of data
  in neutrino and antineutrino modes, respectively. 
 Both experiments operate  
 as a $\nu_\mu \to \nu_\mu/\bar \nu_\mu \to \bar \nu_\mu$
  disappearance experiment as well as a 
   $\nu_\mu \to \nu_e/\bar \nu_\mu \to \bar \nu_e$
 appearance experiment. 
 Their disappearance 
 measurements have no sensitivity to the mass ordering 
 but are the responsible for the 
 precise determination of $\vert \Delta m^2_{32}\vert$ 
 (or equivalently  $\vert \Delta m^2_{31}\vert$).
 T2K results given in \cite{T2K:2023smv} are $\Delta m^2_{32}\vert^{\rm NO} = (2.49\pm 0.05) \times 
 10^{-3}\, \rm eV^2$  and
 $\vert\Delta m^2_{31}\vert^{\rm IO}= (2.46 \pm 0.05) \times 
 10^{-3}\, \rm eV^2$, note that uncertainty is $\sim$ 2\%. 
NOvA's results are given by 
$\Delta m^2_{32}\vert^{\rm NO} = (2.39 \pm 0.06) \times 
 10^{-3}\, \rm eV^2$  and
 $\vert\Delta m^2_{32}\vert^{\rm IO} = (2.44 \pm 0.06)\times 
 10^{-3}\, \rm eV^2$, see \cite{NOvA:2023iam}.  Given the consistency of T2K and NOvA disappearance measurements, they can be combined , as in \cite{Esteban:2020cvm} ,
for both a NO and a IO fit. Using the update of these fits, given in  \cite{NuFIT2022}, we have for NO
\begin{align}
\chi^2_{\rm NO}(\Delta m^2_{31})_{\rm LBL} &=
 \left( \frac{\Delta m^2_{31}-\Delta m^2_{\rm 31} |^{\rm NO} }{ \sigma^{\rm NO}_{\rm LBL} } \right)^2
\end{align}
with  
 $(\Delta m^2_{31}|^{\rm NO}, \sigma^{\rm NO}_{\rm LBL})=(2.516, 0.031) \times 10^{-3
}$ eV$^2$.
Similarly for the  IO fit,
\begin{equation}
\chi^2_{\rm IO}(\vert \Delta m^2_{32}\vert )_{\rm LBL}= \left(\frac{\vert \Delta m^2_{32}\vert -\vert \Delta m^2_{\rm 32}\vert^{ \rm IO} }{\sigma^{\rm IO}_{\rm LBL}} \right)^2
\end{equation}
with $(\vert \Delta m^2_{32}\vert^{\rm IO},\sigma^{\rm IO}_{\rm LBL})=(2.485, 0.031) \times 10^{-3
}$ eV$^2$.  
Note since the uncertainty for both NO and IO are the same, $ \sigma^{\rm IO}_{\rm LBL}=\sigma^{\rm NO}_{\rm LBL}$, we can drop the mass ordering 
on this symbol. Also it is important that this  combined uncertainty is $\sim$ 1.2\%, much less than the RHS of eq. \ref{eq:MOsum} and  \ref{eq:MOsumJ}.
T2K's and NOvA's results are not expected to change significantly in the next few years due to the large statistics already collected by these experiments.
\\

 \textbf{\textit{Combining JUNO with T2K and NOvA}} ---
We combine the results of T2K, NOvA and JUNO by just adding the $\chi^2$ for the combine LBL results to that of JUNO, using the best fit for NO and the measurement precision of JUNO as variables as follows:
\begin{align}
\chi^2_{\rm NO}(\Delta m^2_{31}, \sigma_{\rm JU}) &= \chi^2_{\rm NO} (\Delta m^2_{31}, \sigma_{\rm JU})_{\rm JU}+ \chi^2_{\rm NO}(\Delta m^2_{31})_{\rm LBL} \notag  \\
\chi^2_{\rm IO}(|\Delta m^2_{32}|, \sigma_{\rm JU})& = \chi^2_{\rm IO} (|\Delta m^2_{32}|, \sigma_{\rm JU})_{\rm JU}+ \chi^2_{\rm IO}(\vert \Delta m^2_{32}\vert)_{\rm LBL} \notag 
\end{align}
where for JUNO we use the relationship given in eq. \ref{eq:dmsq32IO31NO} for the best fit values. 
Then the difference in the $\Delta \chi^2$ is given by
\begin{align}
\Delta \chi^2(\Delta m^2_{31}|^{ \rm NO} _{\rm JU}, \sigma_{\rm JU})_{\rm min}  & = \chi^2_{\rm IO}|_{\rm min} -\chi^2_{\rm NO}|_{\rm min}   \,.
\end{align}
Now everything is determined except for the best fit value $\Delta m^2_{31}|^{\rm NO} $ and precision of this measurement $ \sigma_{\rm JU}$ by the JUNO experiment.  In Fig. \ref{fig:JUNOplot} we have plotted the 
 $\Delta \chi^2_{\rm min}$ defined as the minimum
 for IO minus the minimum for NO as a function of  JUNO's $\Delta m^2_{31}\vert^{\rm NO} $ and the fractional precision of the measurement by JUNO\footnote{ We fix the solar parameters at their best fit values determined by 
        NuFIT~\cite{Esteban:2020cvm} $\Delta m^2_{21}= 7.41 \times 10^{-5}$ eV$^2$  and 
        $\sin^2\theta_{12}=0.303$. Small changes in these parameters when measured by JUNO will not effect our results as the experimental uncertainty on $\Delta m^2_{\rm atm}$ from the LBL experiments will dominate. }. 
 On this fan shaped figure we have shown for reference  Daya Bay's result as well as the one from the combined fit of Daya Bay and RENO data given by the NuFIT group.  Clearly as JUNO's precision on the measurement gets better and better one moves down this plot which increases the $\Delta \chi^2$, however the central value could also move right (left) thereby increasing (decreasing) the $\Delta \chi^2$.  This plot assumes that the disappearance $\Delta m^2_{\rm atm}$ results from T2K and NOvA will not  significantly change from what they are now. So, as soon as JUNO presents results on  the $\vert \Delta m^2_{\rm atm}\vert$ fit to the spectrum at the far detector, one can read off from this plot the contribution that   $\nu_\mu/\bar \nu_\mu$ and $\bar{\nu}_e$  disappearance measurements from T2K, NOvA and JUNO make to the  $\Delta \chi^2$ for the mass ordering determination.  If JUNO has the same central value as the NuFIT (React. Comb.)  value but with a precision below 1\%, then the contribution from these disappearance measurements will be greater than 9 units of  $\Delta \chi^2$ , i.e. greater than 3$\sigma$. \\

 \textbf{\textit{Conclusions}} --- We  re-examine an idea we have had on how to determine the neutrino mass ordering, almost 20 years ago,  
 using only neutrino disappearance data in vacuum.  This is auspicious today in the light of 
 the current precision  on $\vert \Delta m^2_{\rm atm}\vert$  (31 for NO and 32 for IO) achieved  by long-baseline experiments ($\sim$2\% individually and  $\sim$1.2\% combined) 
  and of the imminent few per mil determination of  
  $\vert \Delta m^2_{\rm atm}\vert$ by JUNO. This kind of unprecedented accuracy allows one to discuss a {\em mass ordering sum rule for the neutrino disappearance channels} (see eq. \ref{eq:MOsumJ}) which may be used to determine the mass ordering in the near future solely using data from these disappearance experiments. To show this with more clarity, we 
combined in a $\chi^2$ function the present results from T2K and NOvA $\nu_\mu$ and $\bar\nu_\mu$ disappearance measurements with that expected from $\bar \nu_e$ disappearance at  JUNO, as a function of the assumed best fit value ($\Delta m^2_{31}|^{\rm NO}_{\rm JU}$)  and fractional accuracy of the JUNO measurement ($\sigma_{\rm JU}$).
In Fig.~\ref{fig:JUNOplot}  one can see the main result,  the values of $\Delta \chi^2_{\rm min} =\chi^2_{\rm IO}\vert_{\rm min}- \chi^2_{\rm NO}\vert_{\rm min}$ in the plane of JUNO's
$(\Delta m^2_{31}|^{\rm NO} \otimes \sigma )$. This is a very useful figure because as soon as JUNO presents its first result on  their  $\vert \Delta m^2_{\rm atm}\vert$  measurement,
one can read from it if NO is preferred, and if so, how much it is preferred in terms of how many units of  $\Delta \chi^2$. In this manner it is conceivable, if JUNO measures 
 $\vert \Delta m^2_{\rm atm}\vert$  close to the one given by  combining Daya Bay and RENO data, that  NO could be soon (in a year or so)  determined, by the combined (T2K, NOvA and JUNO) disappearance measurements alone,  to better than  3 $\sigma$ i.e. a confidence level of 99.73\%. \\[3mm]

\begin{acknowledgments}

SJP would like to thank Peter Denton for many useful discussions on matter effects for $\nu_\mu$ disappearance.
SJP acknowledges support by the United States Department of Energy under Grant Contract No.~DE-AC02-07CH11359.  
R. Z. F. is partially supported by Funda\c{c}\~ao de Amparo \`a Pesquisa do Estado de S\~ao Paulo (FAPESP) under Contract No. 2019/04837-9, and Conselho Nacional de Desenvolvimento Cient\'ifico e Tecnol\'ogico (CNPq).
We would both like to thank our long time collaborator, Hiroshi Nunokawa, for many enlightening discussions on this topic over many years. Its a pity he chose not  to join us for this follow on paper of \cite{Nunokawa:2005nx}.

This project has also received support from the European Union's Horizon 2020 research and innovation programme under the Marie  Sklodowska-Curie grant agreement No 860881-HIDDeN as well as under the Marie Skłodowska-Curie Staff Exchange grant agreement No 101086085 - ASYMMETRY.
\end{acknowledgments}

\bibliography{NPZ++}% Produces the bibliography via BibTeX.

\appendix

\begin{widetext}
\section{Supplemental Material}

Here we derive the {\it ``mass ordering sum rule for neutrino disappearance channels''}  from the observations of \cite{Nunokawa:2005nx}.
Daya Bay and Reno measure the same $|\Delta m^2_{ee}|$ for both mass orderings.
Therefore from eq. \ref{eq:dmsqee} we have the following:
\begin{align}
 \Delta m^2_{31}|^{\rm NO}_{e ~\rm disp} &= |\Delta m^2_{ee}| + \sin^2 \theta_{12} \Delta m^2_{21} \notag \\
|\Delta m^2_{32}|^{\rm IO}_{e ~\rm disp} &= |\Delta m^2_{ee}| + \cos^2 \theta_{12} \Delta m^2_{21} \notag  \\
|\Delta m^2_{32}|^{\rm IO}_{e ~\rm disp}- \Delta m^2_{31}|^{\rm NO}_{e ~\rm disp} &= \cos 2 \theta_{12} \Delta m^2_{21}.
 \label{eq:Ddmsqee}
\end{align}
Daya Bay's reported measurements given in  \cite{DayaBay:2022orm}  of  
$\Delta m^2_{32}|^{\rm NO}_{\rm DB}=2.466\pm 0.060 \times 10^{-3}$ eV$^2$ and 
$|\Delta m^2_{32}|^{\rm IO}_{\rm DB}=2.571\pm 0.060 \times 10^{-3}$ eV$^2$ satisfy eq. \ref{eq:Ddmsqee}, after adding $\Delta m^2_{21}$ to  $\Delta m^2_{32}|^{\rm NO}$ to obtain $\Delta m^2_{31}|^{\rm NO}$. The agreement is much, much smaller than the quoted uncertainty, demonstrating that these two results are connected via  $|\Delta m^2_{ee}|$.\\

Similarly for the disappearance channels in T2K and NOvA, they both measure  the same $|\Delta m^2_{\mu \mu}|$ for both mass orderings
and matter effects are negligible in this channel for both T2K and NOvA, see \cite{Denton:2024thm}. Therefore from eq. \ref{eq:dmsqmm} we have the following:\\
\begin{align}
 \Delta m^2_{31}|^{\rm NO}_{\mu ~\rm disp}  &= |\Delta m^2_{\mu \mu}| + (\cos^2 \theta_{12}-\sin \theta_{13} \cos \delta^{\rm NO}) \Delta m^2_{21}  \notag \\
 |\Delta m^2_{32}|^{\rm IO}_{\mu ~\rm disp}&= |\Delta m^2_{\mu \mu}| + (\sin^2 \theta_{12}+\sin \theta_{13} \cos \delta^{\rm IO})  \Delta m^2_{21} \notag \\
\Delta m^2_{31}|^{\rm NO}_{\mu ~\rm disp} - |\Delta m^2_{32}|^{\rm IO}_{\mu ~\rm disp} &= (\cos 2 \theta_{12} -2\sin \theta_{13}\, \widehat{\cos \delta} ) \Delta m^2_{21}.  \label{eq:Ddmsqmm}
\end{align}
where $\widehat{\cos \delta} \equiv \frac1{2}(\cos \delta^{\rm NO}+\cos \delta^{\rm IO})$.  T2K provides enough significant figures such that again the identity eq. \ref{eq:Ddmsqmm}  can be checked.
T2K results given in Table 13 of \cite{T2K:2023smv} are reported as $ \Delta m^2_{32}|^{\rm NO}_{\rm T2K} = 2.494 \pm 0.054 \times 10^{-3}$ eV$^2$ and  $ |\Delta m^2_{31}|^{\rm IO}_{\rm T2K} = 2.463 \pm 0.049 \times 10^{-3}$ eV$^2$ again after correcting for the $ 31 \leftrightarrow 32$ for both mass ordering, we get excellent agreement when using $ \widehat{\cos \delta} \approx 0$.  The agreement is much, much smaller than the quoted uncertainty, demonstrating that these two results are connected via  $|\Delta m^2_{\mu \mu}|$. NOvA, unfortunately, does not provide enough significant figures for their measurements to demonstrate the  $|\Delta m^2_{\mu \mu}|$  connection as convincingly as T2K, but their measurements are in agreement with \ref{eq:Ddmsqmm}.\\

Now, adding and rearranging eq. \ref{eq:Ddmsqee} and \ref{eq:Ddmsqmm} we have the {\it ``mass ordering sum rule for neutrino disappearance channels''} given in eq. \ref{eq:MOsum}: 
\begin{align}
( \Delta m^2_{31}|^{\rm NO}_{\mu ~\rm disp} - \Delta m^2_{31}|^{\rm NO}_{e ~\rm disp}) 
~~ + ~~ (| \Delta m^2_{32}|^{\rm IO}_{e ~\rm disp} - |\Delta m^2_{32}|^{\rm IO}_{\mu ~\rm disp}) 
~~  & =  ~~(2 \cos 2 \theta_{12}- 2 \sin \theta_{13} \, \widehat{\cos \delta} )\Delta m^2_{21}\, , \label{eq:MOsum2} \\
& =~~ (2.4 -0.9  \, \widehat{\cos \delta})\% ~ |\Delta m^2_{\rm atm}|  \notag \,,
 \end{align}
 where for the last line we have used $ \Delta m^2_{21}/\vert\Delta m^2_{\rm atm}\vert= 0.03$, $\sin^2\theta_{12}=0.3$ and $\sin\theta_{13}=0.15$.
 Note the interchange between ``$\mu$ disp'' and ``e disp'' when going from NO to IO. Also this sum is invariant if we replace $\Delta m^2_{31}$ with  $\Delta m^2_{32}$ for NO or/and  $\Delta m^2_{32}$ with  $\Delta m^2_{31}$ for IO. We use the  above choice because that is the choice made by NuFIT, but our physics conclusions are independent of this choice. In passing it is worth noting that the  NuFIT results 
 given in \cite{NuFIT2022} suggest that $\widehat{\cos \delta} \leq 0$.\\

Due to the phase advance (NO) or retardation (IO) of the atmospheric oscillations, JUNO does not measure exactly the same  $|\Delta m^2_{ee}|$ for both mass orderings, see \cite{Forero:2021lax}, infact
\begin{align}
 |\Delta m^2_{ee}|^{\rm IO} \approx  1.007 \times  \Delta m^2_{ee}|^{\rm NO} =  \Delta m^2_{ee}|^{\rm NO} + 1.8 \times 10^{-5} {\rm eV}^2.
 \end{align}
This changes eq. \ref{eq:Ddmsqee} and also the {\it ``mass ordering sum rule for neutrino disappearance channels''} to what is given in eq. \ref{eq:MOsumJ} by adding 0.7\, \% times $|\Delta m^2_{\rm atm}|$ to the RHS:
\begin{align}
&( \Delta m^2_{31}|^{\rm NO}_{\rm LBL} - \Delta m^2_{31}|^{\rm NO}_{\rm JU}) 
+ (| \Delta m^2_{32}|^{\rm IO}_{\rm JU} - |\Delta m^2_{32}|^{\rm IO}_{\rm LBL})
 \approx  (3.1 -0.9  \, \widehat{\cos \delta})\% ~ |\Delta m^2_{\rm atm}|  \,. \label{eq:MOsumJ2}
 \end{align}
Here  the label ``e disp'' (`` $\mu$ disap'') has been replaced with the label  ``JU''  (``LBL'') as this sum rule is specific for T2K, NOvA (LBL) and JUNO (JU).   The additional 0.7\,\% has an impact on Fig. \ref{fig:JUNOplot} because of the precision of the JUNO's measurements on $\Delta m^2_{\rm atm}$'s.

If Nature's choice for the mass ordering is NO, then
\begin{align}
( \Delta m^2_{31}|^{\rm NO}_{\rm LBL} - \Delta m^2_{31}|^{\rm NO}_{\rm JU}) \approx 0 \quad {\rm and} \quad (| \Delta m^2_{32}|^{\rm IO}_{\rm JU} - |\Delta m^2_{32}|^{\rm IO}_{\rm LBL}) \approx 
 (3.1 -0.9  \, \widehat{\cos \delta})\% ~ |\Delta m^2_{\rm atm}|
\end{align}
 whereas for IO
\begin{align}
( \Delta m^2_{31}|^{\rm NO}_{\rm LBL} - \Delta m^2_{31}|^{\rm NO}_{\rm JU}) \approx   (3.1 -0.9  \, \widehat{\cos \delta})\% ~ |\Delta m^2_{\rm atm}| \quad {\rm and} \quad (| \Delta m^2_{32}|^{\rm IO}_{\rm JU} - |\Delta m^2_{32}|^{\rm IO}_{\rm LBL}) \approx 
0 \,,
 \end{align}
 where ``$\approx$'' should be interpreted to mean ``within measurement uncertainties''.  The current measurement uncertainty for the combined  T2K and NOvA  measurement is  $\sim$\,1.2\%.  JUNO's measurement uncertainty will reach 0.8\,\% within one year of data taking and reach 0.2\,\% within 6 years.  If T2K and NOvA can improve their measurement uncertainties on the  $\Delta m^2_{\rm atm}$'s even modestly, in the future, this can impact the $\Delta \chi^2$ between the two mass orderings substantially.\\
 
\end{widetext}

\end{document}